\documentclass[]{spie}  
\usepackage[]{graphicx}

\title{Planet Formation Imager (PFI): Introduction and Technical Considerations\footnote{\small~Copyright 2014 Society of Photo-Optical Instrumentation Engineers. One print or electronic copy may be made for personal use only. Systematic reproduction and distribution, duplication of any material in this paper for a fee or for commercial purposes, or modification of the content of the paper are prohibited. DOI abstract: http://dx.doi.org/10.1117/12.2057262}} 

\author{John D. Monnier\supit{a},  Stefan Kraus\supit{b}, David Buscher\supit{c}, J.-P. Berger\supit{d}, Christopher Haniff \supit{c},  Michael Ireland\supit{e},
Lucas Labadie\supit{f},
Sylvestre Lacour\supit{g},
Herv\'{e} Le Coroller\supit{h}
Romain G. Petrov\supit{i},
J\"{o}rg-Uwe Pott\supit{j},
Stephen Ridgway\supit{k}
Jean Surdej\supit{l},
Theo ten Brummelaar\supit{m},
Peter Tuthill\supit{n}, and
Gerard van Belle\supit{o},
\skiplinehalf
\supit{a}University of Michigan, USA; \\
\supit{b}University of Exeter, UK;\\
\supit{c}University of Cambridge, UK;\\
\supit{d}ESO, Germany;\\
\supit{e}Australian National University, Canberra, Australia;\\
\supit{f}University of Cologne, Germany;\\
\supit{g}Observatoire de Paris, France;\\
\supit{h}Laboratoire d'Astrophysique de Marseille, Marseille France;\\
\supit{i}University of Nice - Sophia Antipolis, Nice, France;\\
\supit{j}MPI for Astronomy, Germany;\\
\supit{k}NOAO, USA;\\
\supit{l}University of Li\`ege, Belgium;\\
\supit{m}Georgia State University, USA;\\
\supit{n}University of Sydney, Australia;\\
\supit{o}Lowell Observatory, USA
}

\authorinfo{Further author information: (Send correspondence to J.D.M.)\\J.D.M.: E-mail: monnier@umich.edu}

\begin{document} 
  \maketitle 

\begin{abstract}
Complex non-linear and dynamic processes lie at the heart of the planet formation process. 
Through numerical simulation and basic observational constraints, the basics of planet formation 
are now coming into focus. High resolution imaging at a range of wavelengths will give us a glimpse 
into the past of our own solar system and enable a robust theoretical framework for predicting 
planetary system architectures around a range of stars surrounded by disks with a diversity of 
initial conditions. Only long-baseline interferometry can provide the needed angular resolution 
and wavelength coverage to reach these goals and from here we launch our planning efforts.
The aim of the ``Planet Formation Imager'' (PFI) project is to develop the roadmap for the 
construction of a new near-/mid-infrared interferometric facility that will be optimized to unmask 
all the major stages of planet formation, from initial dust coagulation, gap formation, evolution 
of transition disks, mass accretion onto planetary embryos, and eventual disk dispersal. 
PFI will be able to detect the emission of the cooling, newly-formed planets themselves over the first 100~Myrs, 
opening up both spectral investigations and also providing a vibrant look into the early 
dynamical histories of planetary architectures.  

Here we introduce the Planet Formation Imager (PFI) Project (www.planetformationimager.org) and give initial thoughts on possible facility architectures and technical advances that will be needed to meet the challenging top-level science requirements.
\end{abstract}



\section{INTRODUCTION}
\label{sec:intro}  

The optical and infrared interferometry community has organized a series of meetings over the past decade on the topic of ``The Future of Optical Interferometry.''  At least 7 meetings held around the world (Li\`ege, Tucson, Lisbon, Socorro, Flagstaff, Turku) brought together astronomy topic experts and instrumentalists to imagine how a future long-baseline interferometer might be used to solve critical problems in stellar astronomy and extragalactic science.  These efforts were buoyed by strong endorsements by Astro2010 Decadal Survey (in the USA)  and Astronet (in Europe) of relevant key astronomy areas such as star and planet formation, exoplanets, active galactic nuclei and certain areas of fundamental stellar properties.  These topics specifically  will be revolutionized by the high angular resolution possible only from long-baseline interferometry.  In a post-ELT and post-JWST era, interferometry will be essential to continue our current pace of discovery in astronomy.
While much debate at these meetings centered on how many telescopes at what size aperture would be needed for various science cases, it was clear that funding for any ambitious new facility would be difficult to obtain in the current budget climate.  At these meetings, no clear path forwards emerged except to continue to make science-driven upgrades to current facilities and to better develop the interferometer observer community along with fresh science cases.

Last year, the European Interferometry Initiative (EII) ``Future of European Interferometry" subgroup helped plan another workshop titled ``Improving the performances of current optical interferometers and future design,'' held in September 2013 at the Observatoire de Haute Provence. Through a series of talks, rich dialogue between participants, and a final round table discussion, a new consensus emerged amongst the attendees\cite{surdej2014}.  The community should focus on a {\em single compelling science case} and aggressively investigate the technical requirements needed to build a capable facility.  This approach should lead to a more unified and coherent effort throughout the community to develop the technologies needed to make a next-generation facility feasible.   With a clear set of top-level science requirements, a well-defined and justified technical roadmap can be created to guide priorities.

Many exciting science topics were discussed -- astrometry and characterization of exoplanets, imaging of active galactic nuclei, spectro-interferometry of mass-losing stars -- but the one that resonated most was the concept of directly imaging the key stages and processes involved in planet formation.

\section{Key Science Case for Planet Formation Imager (PFI)} 

The goal of Planet Formation Imager (PFI) is to image all the key processes guiding planet formation around young stars and to follow early dynamical interactions between giant planets within the first 100 million years.  Planet formation is one of the most important fields of astronomy, connecting star formation with exoplanets.  A robust understanding of planet formation is needed to understand the demographics we see of exoplanet architectures around mature stars based on data from conventional planet finding efforts, such as radial velocity and transit surveys.  Numerical hydrodynamicists as well as analytic theorists have taken up the challenge to develop a general theory of planet formation and the subsequent disk migration phenomena, efforts that include chemistry, dust formation and growth, radiative transfer, dust dynamics, magnetic field evolution in disks, viscous accretion disk theory, dynamics, and more.  New high angular resolution facilities will accelerate discoveries, most especially through mm-wave imaging (ALMA), scattered light polarimetric imaging (GPI/SPHERE/SEEDS), high contrast coronagraphy (GPI/SPHERE/Subaru), and mid-IR interferometric imaging with MATISSE/VLTI.   Furthermore, JWST and the ELTs will push the angular resolution of single-aperture techniques by another factor of 3 or so.  Current trends point to a robust future for the fields of planet formation and exoplanet studies.

 \begin{figure}
   \begin{center}
   \begin{tabular}{c}
   \includegraphics[height=7cm]{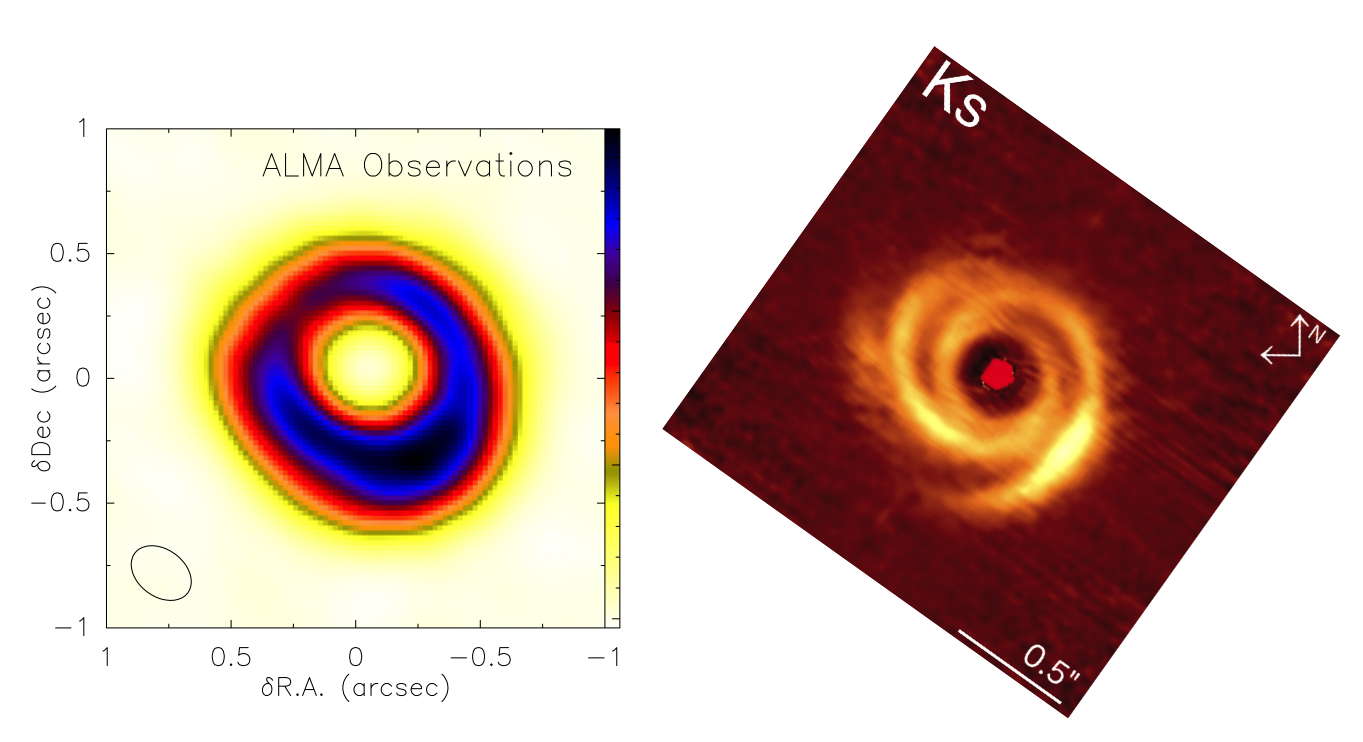}
   \end{tabular}
   \end{center}
   \caption[example] 
   { \label{fig:almanaco} 
(left panel) This ALMA image of thermal emission from large grains around SAO~206462 shows evidence for a lopsided ring (from Perez et al. 2014; Ref. \citenum{perez2014}).  (right panel) NACO/VLT imaging in polarized emission reveals even more details, including spiral structures extending into the inner disk (from Garufi et al. 2013; Ref. \citenum{garufi2013}).  There is a clear need for angular resolution beyond what ALMA and even 30-m telescopes can bring.   The Planet Formation Imager (PFI) will image disks with $<$1 milliarcsecond angular resolution. } \end{figure} 

Figure~\ref{fig:almanaco} shows recent ALMA imaging and VLT/NACO scattered-light polarimetric imaging of Herbig Ae star SAO~206462.  Even the short-baseline ALMA images\cite{perez2014} begin to resolve structure in the dust emission from this object, including a deficit of emission in the center and an asymmetric, elongated ring structure.  VLT/NACO imaging\cite{garufi2013} with $\sim$50~milliarcsecond angular resolution clearly finds spiral arm structure in the polarized scattered-light emission from small grains in the upper layers of the disk.  These disk structures are not straightforward to understand with current theories of planet formation.  ALMA will achieve more than 10$\times$ better angular resolution soon, which will give us a new view into these questions.  

\begin{figure}
   \begin{center}
   \begin{tabular}{c}
   \includegraphics[height=7cm]{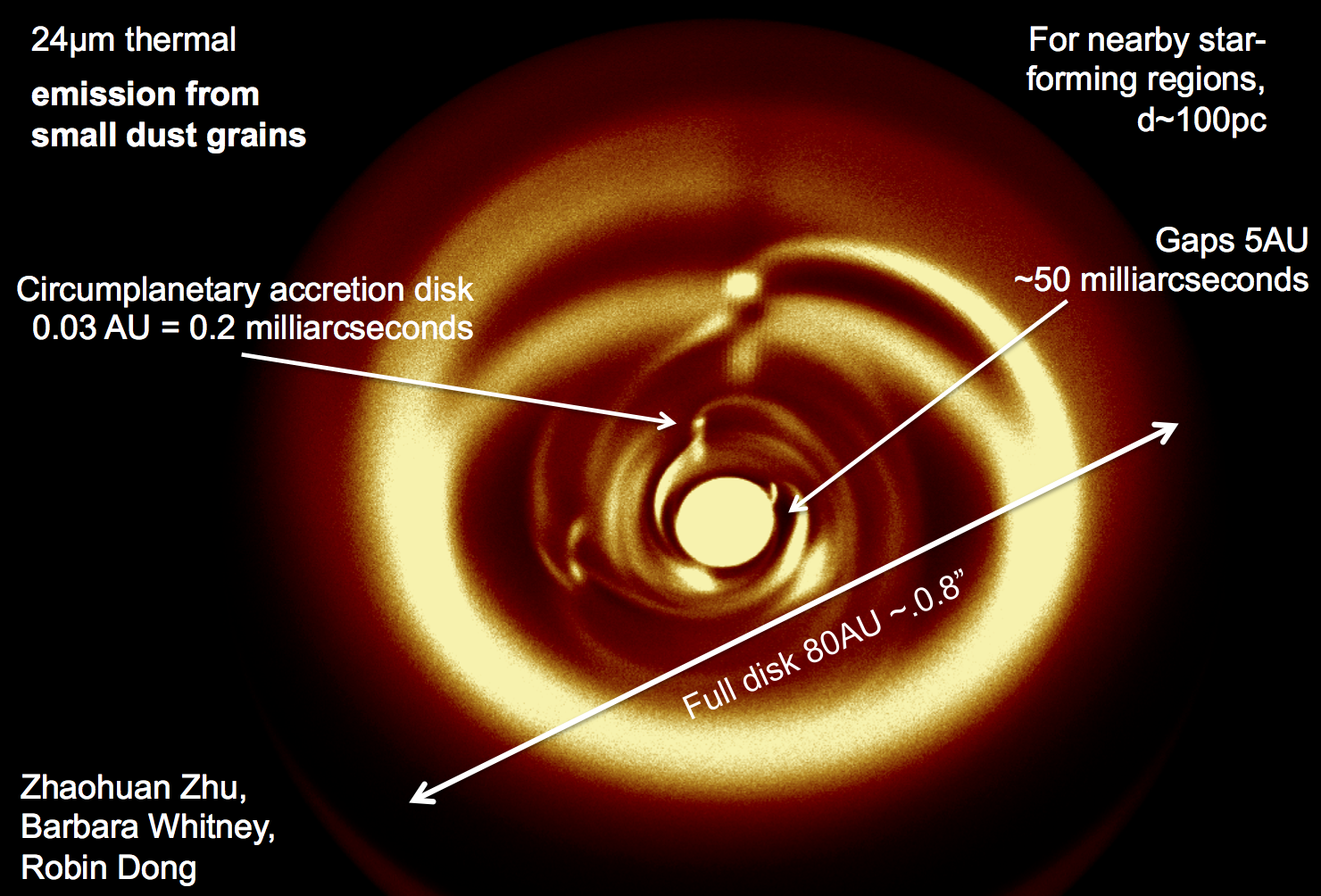}
   \end{tabular}
   \end{center}
   \caption[example] 
   { \label{fig:diskscales} 
Here we see a result from a radiative transfer calculation (Robin Dong, private communication) of a disk undergoing vigorous planet formation (as described by Zhu et al. 2011; Ref. \citenum{zhu2011}).  For nearby star forming regions (d$\sim$100~pc), thermal infrared disk emission might come from regions of size $\sim$80~AU$\sim$ 0.8'', with sub-structures on a range of fine scales.  We might expect gaps of size from 5~AU~$\sim$50~milliarcseconds with circumplanetary accretion disks with size close to the planetary Hill radius $\sim$0.03~AU$\sim$0.2 milliarcseconds.   In order to cover all relevant scales of planet formation, PFI should reach down to sub-milliarcsecond resolution over about a 0.5" field of view with priority given to the thermal infrared.   See additional discussion in Kraus et al. (this volume). } \end{figure}

Unfortunately, neither the full ALMA nor scattered light imaging with even the 30-m ELTs will suffice to resolve the important planet formation physical length scales for lower-mass T Tauri stars.   Figure~\ref{fig:diskscales} shows the results of a radiative transfer simulation of a hydrodynamic calculation of multi-planet formation\cite{zhu2011}.  We see that while current methods can probe the overall disk structure and can resolve large gaps or internal dust cavities, the accretion streams and circumplanetary accretion disks lie on spatial scales orders of magnitude smaller than ever achievable with ALMA or large single-aperture telescopes.  We must look to a new project to probe planet formation on the actual scales that regulate planet formation.   

The PFI Science Case is broader than only planet formation studies through imaging dust emission.  Please see the accompanying SPIE paper by Stefan Kraus (in this volume) for further details on the science cases for PFI, including  how PFI will image the dynamical relaxation of giant planet orbital configurations over the first 100 Myrs.

\section{PFI Project Organization} 
\label{sec:title}
The Planet Formation Imager (PFI) project was initiated in November 2013 by John Monnier, Jean-Philippe Berger, and Stefan Kraus, in conjunction with a Kick-off committee consisting of Mike Ireland, Lucas Labadie, Sylvestre Lacour, Jorg-Uwe Pott, Steve Ridgway, Jean Surdej, Theo ten Brummelaar, Peter Tuthill, and Gerard Van Belle (with recent additions of Romain Petrov and Chris Haniff).  The goal of the PFI Project is currently to:

\begin{enumerate}
\item Formulate the science requirements and identify the key enabling technologies;
\item Build support in the broad astronomical community as well as the interferometry community;
\item Start lobbying with decision makers (ASTRONET, ESO, Decadal Review);
\item Prepare for upcoming funding opportunities (OPTICON, MSIPS)
\end{enumerate}

The PFI project has established a set of 3 executive officers and held internal elections.  John Monnier was elected as Project Director, Stefan Kraus as Project Scientist, and David Buscher as Project Architect.  The Project Director is responsible for overall organization of PFI and is the main point of contact for the Project.  S/he will hold regular meetings with the advisory board (PAC, Project Advisory Committee), will lead regular telecom meetings, and be responsible for public outreach through emails, websites, and social media.  The Project Scientist will lead the Science Working Group (SWG) to develop and prioritize key achievable science cases, including producing and maintaining the ``Top Level Science Requirements (TLSR)''.  The Project Architect will lead the Technical Working Group (TWG) to conduct concept studies in order to identify key technologies and to develop a technology roadmap consistent with the TLSRs.  

In May 2014, we published a public ``Call for Participation in PFI Concept Studies."  It read as follows:
\begin{quote}
We are happy to announce the First Call for Participation in PFI Concept Studies.  The ambitious goal of PFI is to image planet-forming disks in nearby star-forming regions with high enough spatial resolution to resolve the key physical processes at work, to witness planet formation live as it happens with $\sim$ 0.1AU resolution or better. Scientists from more than a dozen different institutes in six countries have begun planning for initial Concept Studies, an effort led by Project Director John Monnier (U. Michigan), Project Scientist Stefan Kraus (U. Exeter), and Project Architect David Buscher (U. Cambridge). Our top priorities for the next 12-24 months will be to define the most compelling areas of science to drive the instrument concept and at the same time determine feasible architectures for meeting the science goals. We seek contributions from the international astronomical community and invite participants to join the PFI Science Working Group or the Technical Working Group. For more information and to sign up to participate, please see http://www.planetformationimager.org (initial deadline to join working groups is June 16, 2014).
\end{quote}

At the time of this writing, over 60 volunteers have expressed interest in joining the TWG and a similar number for the SWG.  We are currently in the process of organizing subgroup leaders and will begin Project work during the Summer 2014.

One open issue is the timescale for building PFI. We hope to clarify this issue within 12-24 months following the work of the Science and Technical Working Groups.  Realistically, most of us involved in PFI recognize that while the ambitious goals of PFI are definitely achievable, they will not be easy to meet with current technologies at a modest price point. However, we are confident that with a clear and strong design reference mission and a well-defined technology roadmap, a feasible timescale for PFI will be one that envisions deployment sometime after the Extremely Large Telescope (ELT) projects are fully operational in the late 2020s.

\section{Technical Considerations}

Alongside this article is a PFI Science Case paper  by Kraus et al. that gives an overview of the priority science cases and the chief considerations guiding the top-level science requirements.  Here, we will focus on technical considerations for PFI, including initial thoughts on the core science requirements, possible facility architectures, site considerations, and some crucial technologies.

\subsection{Top-Level Science Requirements}
In order to guide our architecture discussion, we must adopt a preliminary set of top-level science requirements.  Based on initial work by the PFI Project, we adopt the following TLSRs for the purposes of this article:

\begin{itemize}
\item The PFI angular resolution is set to resolve the region around an embedded giant exoplanet in the disk where the planet dominates the gravitational influences, the so-called ``Hill Sphere.''  For a one Jupiter-mass planet at 1 AU around a sun-like star, we would require multiple resolution elements across 0.1~AU.  We adopt a requirement of 0.03~AU resolution for distances up to 140pc (distance to Taurus star forming region), thus PFI requires 0.2 milli-arcsecond resolution.
\item PFI should be sensitive to thermal dust emission over a range of temperatures to probe the region where giant planets are thought to form and to eventually migrate.  Here, we focus on the mid-infrared ($\lambda\sim10\mu$m) where dust grains at 300K are peaking in their thermal emission, giving us good sensitivity to the terrestrial planet forming zone.
\item In order to have 0.2 mas resolution at 10$\mu$m, PFI will require 10km baselines.
\item PFI should be sensitive enough to observe a range of structures in the disk. For instance (for d$=$100pc):
\begin{itemize}
\item T Tauri star photosphere: N(mag)$=$7.5
\item Best case circumplanetary disk (a la LkCa15 object\cite{kraus2012}): N(mag)$=$11
\item 10~Myr old Jupiter: N(mag)$=$15.7
\item 100~MYr old Jupiter: N(mag)$=$18.5
\item For reference, the current VLTI/MIDI with fringe tracking on the UTs can observe N(mag)$=$7.0
\end{itemize}
\item Lastly, PFI will need excellent imaging capabilities since we expect disk emission to be complicated (as shown by ALMA and scattered light imaging for Herbig Ae/Be stars). If we aim for images with 400$\times$400 pixel imaging, then we must have similar imaging to the radio VLA. PFI will likely need 20--30 telescopes in the array.
\end{itemize}

\subsection{Architecture Overview}

We consider a number of architectures below that could satisfy the above TLSRs. Note that many of the conclusions below are highly preliminary and will be critically scrutinized by the SWG and TWG over the coming 12-24 months using facility simulations.  It is beyond the scope of this brief overview paper to justify all the claims below and we include specific numbers here just to give a snapshot of the PFI Project's current thinking before the detailed studies begin in earnest later in 2014.

\begin{enumerate}
\item {\bf Conventional ground-based interferometer array.} This system will focus on mid-infrared wavelengths (5--13$\mu$m) with 10 km baselines.  This will require multi-km vacuum pipes with diameter $>$0.5m to minimize fresnel diffraction.  For conventional sites and using $D<10$m telescopes, N-band fringe tracking on the $N=7.5$ central star will not be possible.  We must use a near-infrared fringe tracker to allow coherent integration in the N-band, similar to what is now done for VLTI/UTs\cite{muller2014}.  Natural guide star adaptive optics will be sufficient on $\sim$4-m class telescopes to allow both fringe tracking and mid-IR sensitivity, assuming $\sim$30~sec coherent integrations are possible.  For imaging complex scenes and to have sufficient short baselines for bootstrapping, we will need $N>20$ telescopes.  The $\sim$30~sec coherent integrations will allow the star to be detected at 10$\mu$m, but much longer bispectrum integrations will be needed to reach $N=15$ -- approximately 20 hours for twenty 4-m telescopes. This will allow individual giant planets to be detected around stars up to 10 Myrs. Note that this architecture could allow near-infrared imaging with  $\sim$50~{\em micro}-arcsecond angular resolution down to H(mag)$=$14 in principle, supporting robust ``ancillary'' science cases in stellar astrophysics, AGN, microlensing, and more.  Future detailed simulations will determine the minimum-sized aperture required for our core science -- we emphasize the above SNR calculation for 4-m apertures is merely illustrative.

\item {\bf Heterodyne array.} This system would be a mid-infrared heterodyne system similar to an expanded Infrared Spatial Interferometer (ISI)\cite{hale2000}.  While the ISI was limited to very bright sources [N(mag)$<$0], new technological advances\cite{lecoarer2014} could drastically improve sensitivity as a possible PFI architecture.  Ireland et al. (this volume) discuss how advances in frequency combs could allow the entire mid-IR bandwidth to be used, an increase over the ISI bandwidth of $>$1000.   The heterodyne detection process at 10$\mu$m is equivalent to observing with a thermal background of 1400K, a factor many times higher than the 273K thermal background that a direct detection scheme suffers.  However, the heterodyne detection process can be done closer to the telescopes with far fewer reflections, thus obtaining a transmission bonus.  In addition, the heterodyne system allows amplifications of the RF signal following detection without introducing additional noise, thus all telescope combinations could be correlated without loss of SNR -- an advantage over conventional ``direct detection'' schemes.   The heterodyne array will still require near-infrared fringe tracking using AO-corrected 4-m class telescopes in order to allow coherent integration in the mid-IR.  While the mid-IR delays can be done in a computer following frequency down conversion (without expensive large vacuum pipe transport), conventional delay lines will still be needed for the near-IR fringe tracking channel (although perhaps only for relatively short bootstrapping baselines).  The balance of advantages and disadvantages will have to be carefully considered by the Technical Working Group.  One exciting possibility is that fiber optics could be used to transport near-infrared light from the telescopes to the beam combining lab which would obviate the need for long vacuum pipes connecting the telescopes to the lab.  Please see the contribution by Ireland et al. (this volume) for detailed considerations on the intriguing possibility of a mid-IR Heterodyne interferometry revival.
 
 \item {\bf Space interferometer.} The mid-infrared thermal background in space is about {\em 26 million times} lower than from the ground.  It is impressive to consider that a 1mm aperture cooled space telescope is {\em in principle} as sensitive as a 8m ground-based telescope for point source detection in the thermal IR.  A space interferometer will need 10km baselines just as the other designs, thus requiring long-distance formation flying capabilities not dissimilar from those being developed for the gravity wave interferometer LISA and with some technology connections with the formation flying maneuvers over short distances that are now routine for  remote docking of cargo ships to the International Space Station (see also Ref. \citenum{lecoroller2012}). One huge advantage of a space platform is that PFI would have access to imaging across the thermal IR, from the mid- to far-IR. 
 This would open up new science cases unthinkable from the ground, including studies of ices, mineralogy, debris disks, etc.    It is encouraging that the NASA's 2013 Strategic 
 Astrophysics Roadmap highlighted space interferometry 
 as an essential technology in the ``Visionary Era'' in optical, IR, and even X-ray astronomy. 
   While much work has been done studying infrared space interferometry in the context of detecting terrestrial planets (TPFI\cite{lay2007}, Darwin\cite{leger2007}), the sensitivity and nulling requirements for PFI are more relaxed than for these earlier planning efforts. Indeed, while imaging disks will be easier in some ways compared to imaging Earth-like planets, space-PFI will require many more telescopes and reconfigurations than TPF/Darwin to allow high fidelity imaging of complex scenes.  Space-PFI may have more in common with the ``Stellar Imager\cite{carpenter2003}'' project than Terrestrial Planet Finder. Thinking ahead, space-PFI will be a natural follow-on to JWST and will  afford a robust extragalactic science case in addition to the Planet Formation Imaging science.  We note that the far-IR space community is already looking into detailed studies of space interferometers (FISICA, SPIRIT/SPECS\cite{leisawitz2008}), although generally with far fewer telescopes and over modest $B<100$m baselines. We share many technology requirements and hope to coordinate our efforts to invest in space interferometry technology studies.
 
 \item {\bf ALMA with longer baselines.}  ALMA is already beginning to do breakthrough imaging of dust continuum and molecules in disks.  Eventually ALMA will have 10-20$\times$ greater angular resolution when the 350$\mu$m band observations are extended to the full 16km baselines, yielding an angular resolution of $\sim$5 milliarcseconds.  This will allow AU-sized structures to be imaged in nearby star-forming regions, an incredibly exciting prospect for learning about planet formation.  While there is no more space on the high plateau site, it might be possible to extend ALMA baselines by connecting with new telescopes built on nearby plateaus to boost resolution to the sub-milliarcsecond level. There is some effort to identify possible locations (e.g., the Long Latin America Millimeter Array LLAMA\cite{arnal2011}).  The advantage of extending an existing successful facility is manifest although the science potential may not be as great as for a shorter wavelength facility. For instance, it might not be possible to detect exoplanets themselves in the mm/sub-mm and also the thermal emission from small grains will be largely invisible in the sub-mm.  It is clear that a mid-IR PFI would be a powerful complement to the mm-wave ALMA and the Science Working Group will look into scientific trade-offs of observing at different wavelengths.
 
 \item{\bf Ground-based coronagraph.}  The ``Planet Formation Imager'' moniker was first attached to a 2nd generation Thirty-Meter Telescope instrument concept\cite{vasisht2006}.  Indeed, a visible-light extreme AO on a 30-m telescope will have $\sim$4~milliarcsecond resolution, similar to sub-mm ALMA.  However, for the same reasons as above, 4~mas is not sufficient to resolve a 1-Hill-Sphere gap in a young disk or to isolate a circumplanetary disk from its surrounding material.  That said, an ELT-PFI will be a powerful complement to mid-IR PFI by probing scattered light that is sensitive to small dust particles in the upper layers of the disk. As for ALMA, the SWG will scrutinize the trade-offs between the baseline PFI concept of a mid-IR, sub-mas resolution array and a 30-m coronagraph instrument.  Furthermore, a 3--5$\mu$m imaging system on an ELT might be sufficient to detect circumplanetary accretion disks and young exoplanets, either complementing PFI or perhaps moving PFI away from this science toward the more challenging (and rewarding) thermal IR dust imaging core science case. 
 
 \item{\bf Space occulter.} Advances in starshade technology in the context of exoplanet discovery\cite{kasdin2012} 
 suggest new possibilities for space-based imaging systems.  Angular resolution can be obtained through occultations of astronomical bodies by large shades in a way similar to classic lunar occultation work\cite{ridgway1977}.  Fresnel interference sets the resolution to be proportional to $\theta\propto\sqrt{\frac{\lambda}{D}}$, which unfortunately means that a 30AU separation is required between a 10km occulting body and the distant collecting telescope to obtain milli-arcsecond resolution at $\lambda=$10$\mu$m.  This does not seem a feasible prospect, even using asteroids as natural occulters\cite{dunham2013}.
 
 \end{enumerate}

\subsection{Sites}

Optical and infrared interferometers typically require the very best sites although adaptive optics and fringe tracking on relative bright targets might relax these conditions.  In addition to considering conventional mid-latitude sites, the PFI project will seriously investigate the unusual attributes of the Antarctic High Plateau Sites (Dome A, C, F) which offer immense promise for both thermal infrared observing (low background) and excellent atmospheric stability\cite{lawrence2004} that allow for much longer coherence times and might allow for novel off-axis fringe tracking.  Some aspects of the PFI core science might be done in Antarctica with a specialized few-telescope system, for instance searching for hot exoplanets and accreting circumplanetary disks in the 3--5$\mu$m range -- similar to the ELT-PFI case discussed in the last section.

The core science of PFI requires a large number of young stars with disks for study. This limits us to considerations of the nearest young associations with active planet formation.  Any site selection will need to account for the accessibility to the well-known Taurus, Orion, and Ophiuchus regions.  In addition, PFI would like to study the evolving characteristics of  cooling gas giant planets in stars found in young clusters over a range of ages ($<$100~Myrs).  The SWG will collect a comprehensive database of nearby star-forming regions to guide these deliberations.

\subsection{Technology Spotlights}
The Technical Working Group will also identify key technologies that could prove decisive to making PFI feasible.  
Although we still have to confirm that a 10-km array of 20--30 telescopes with 4 to 8\,m apertures will give the required sensitivity, angular resolution, dynamic range, and sensitivity to dynamics, there are no known technology showstoppers to PFI.  Many believe we could build a conventional ground-based interferometer version of PFI with today's technology (e.g., Buscher et al, these proceedings).   If PFI can not be built until after the ELTs, then there is time to invest in new technologies that might significantly reduce the cost and risk of a PFI facility -- these efforts may prove crucial.

Here, we highlight a few technologies that could have a large influence on the direction of PFI moving forward.

\begin{itemize}
\item {\bf Inexpensive 4-m to 8-m class telescopes.}  While there is some skepticism within our Project that major advances are possible in this area, we believe the community should focus on developing breakthroughs in this area.  PFI will greatly benefit from large numbers of telescopes and thus unit costs should be minimized.  There have been advances in carbon fiber reinforced polymer mirrors\cite{andrews2010} that, when coupled with inexpensive low-order adaptive optics, could radically change our telescope designs.  For PFI, the mirror specifications can be degraded  compared to  typical optical systems -- PFI will focus on the infrared only, the telescopes will naturally incorporate adaptive optics, and they also will require only a narrow field-of-view.  Scattered light off high-order surface imperfections might reduce throughput but will not ruin the imaging ability for a single-mode system like PFI.   There were new ideas presented in 3d printing, active optics, and more on display at this meeting in Montreal.  Current interferometer groups might consider partnering with industry to field-test these technologies over the coming decade.

\item {\bf Frequency combs.} The heterodyne concept will require mid-IR frequency combs that do not yet exist.
\item{\bf Fiber optics.} The use of single mode fibers for beam transport might be crucial for the heterodyne architecture. Much work has been done in this area for the OHANA\cite{perrin2003} project and new lab results were recently presented at this meeting (Anderson et al.).  
\item {\bf Mid-IR photonics.}  For the direct detection scheme, we will need efficient mid-IR combiners that can combine large numbers of baselines simultaneously.  This work is just beginning with new materials and some progress was reported at this meeting (see, e.g., Cvetojevic et al., Martiarena et al., Minardi et al. )
\item{\bf Space technology.}  Any progress towards space interferometry technology demonstrations is welcome and sorely needed.  There is renewed interest in nano- or micro-sat flight opportunities to test ideas in formation flying, fiber nulling combination (Lacour et al., this meeting), and more.  
\end{itemize}

Other interesting technologies include inexpensive laser guide stars, realistic 10km delay lines (super low-loss multiple reflections from LIGO legacy, ``ultra-flat'' zero dispersion fiber optics), fringe tracking (demonstrate 30 second coherencing, perform multi-hour bispectrum integrations), novel combiners for large numbers (N$>$20) of beams (both near-IR and mid-IR, densified pupil coronagraph), 40$\mu$m heterodyne detectors for dome A (could avoid near-IR fringe tracking possibly), new operations models (take advantage of robotic telescopes and large redundancy to slash operating costs).

\subsection{Intermediate Milestones on the Roadmap}

There are possible intermediate facilities that might demonstrate PFI technologies while advancing PFI science  goals.  

\begin{itemize}
\item L/M band imaging using J/H/K band coherencing:  VLTI-MATISSE at 5--10$\mu$m (this project is funded with results expected by 2020), L/M band modes for CHARA/MROI (not funded at the moment)
\item Imaging NIR scattered-light disks in polarized emission with ``ALMA resolution'' (5~mas): NIR polarization+nulling instrument at VLTI -- a kind of high-resolution GPI/SPHERE.
\item Antarctica High Plateau interferometer: 1--5$\mu$m optimized for exoplanets.
\end{itemize}

\section{Conclusions}

While some argue that ``interferometry is inevitable'' (see Rinehart paper, these proceedings), there is no guarantee that the prodigious pace of discovery we currently enjoy in Astronomy is sustainable.   The PFI Project hopes to identify and specify compelling science goals to guide our thinking into the post-ELT era.  This is the first SPIE meeting for PFI and we hope to see maturing of the concept over the next few years.  PFI is a long-term goal requiring coherent efforts over many years, possibly decades, to reach fruition.  We welcome new participation and encourage interested readers to visit {\bf planetformationimager.org} to get involved in one of the working groups.


\acknowledgments     
The PFI Project wishes to thank the organizers of the 2013 workshop at the Observatoire de Haute Provence (OHP), where PFI was born. JDM especially wishes to think EII for travel support that allowed him to attend the OHP meeting.


\bibliography{monnier_PFI}   
\bibliographystyle{spiebib}   

\end{document}